\documentclass[twocolumn,abs,table,dvipsnames]{aastex62}

\begin{document}
\title{The Clearing Timescale for Infrared-selected Star Clusters in M83 with HST}

\author[0009-0007-9657-1412]{Suyash Deshmukh}
\affiliation{Department of Astronomy, University of Massachusetts at Amherst, Amherst, MA 01003, USA}

\author[0000-0002-1000-6081]{Sean T. Linden}
\affiliation{Steward Observatory, University of Arizona, 933 N Cherry Avenue, Tucson, AZ 85721, USA}

\author[0000-0002-5189-8004]{Daniela Calzetti}
\affiliation{Department of Astronomy, University of Massachusetts at Amherst, Amherst, MA 01003, USA}

\author[0000-0002-8192-8091]{Angela Adamo}
\affiliation{Department of Astronomy, Oskar Klein Centre, Stockholm University, AlbaNova University Centre, 106 91 Stockholm, Sweden}

\author[0000-0003-1427-2456]{Matteo Messa}
\affiliation{INAF – OAS, Osservatorio di Astrofisica e Scienza dello Spazio di Bologna, via Gobetti 93/3, I-40129 Bologna, Italy}

\author[0000-0002-3247-5321]{Kathryn Grasha}
\affiliation{Research School of Astronomy and Astrophysics, Australian National University, Canberra, ACT 2611, Australia}

\author[0000-0003-2954-7643]{Elena Sabbi}
\affiliation{Space Telescope Science Institute, 3700 San Martin Drive, Baltimore, MD, USA}

\author[0000-0002-0806-168X]{Linda Smith}
\affiliation{European Space Agency (ESA), ESA Office, Space Telescope Science Institute, 3700 San Martin Drive, Baltimore, MD 21218, USA}

\author[0000-0001-8348-2671]{Kelsey E. Johnson} 
\affiliation{Department of Astronomy, University of Virginia, Charlottesville, VA, USA}

\begin{abstract}
We present an analysis of Hubble Space Telescope (HST) data from WFC3/UVIS, WFC3/IR and ACS, investigating the young stellar cluster (YSC) population in the face-on spiral galaxy M83. Within the field of view of the IR pointings, we identify 454 sources with compact F814W continuum and Pa$\beta$ line emission with a S/N $\geq 3$ as possible YSC candidates embedded in dust. We refine this selection to 97 candidates based on their spectral energy distributions, multi-wavelength morphology, and photometric uncertainties. For sources that are detected in all bands and have mass $> 10^{2.8} M_{\odot}$ (53 sources), we find that by 2 Myr $75\%$ of infrared-selected star clusters have an $A_{V} \leq 1$, and that by 3 Myr the fraction rises to $\sim 82\%$. This evidence of early clearing implies that pre-supernovae feedback from massive stars are responsible for clearing the majority of the natal gas and dust that surround infrared-selected star clusters in M83. Further, this result is consistent with previous estimates based on WFC3 observations, and adds to the growing body of literature suggesting pre-supernova feedback to be crucial for YSC emergence in normal star-forming galaxies. Finally, we find a weak correlation between the YSC concentration index and age over the first 10 Myr, which matches previous studies and indicates little or no change in the size of YSCs in M83 during their early evolution.

\end{abstract}

\keywords{galaxies: star clusters: general - galaxies: spiral - galaxies: ISM - galaxies: M83}

\section{Introduction}
Given that the majority of stars form within stellar clusters \citep[][]{lada_embedded_2003}, it is crucial to understand the formation and evolution of these systems. This paper focuses specifically on the early stages of cluster formation, and in particular the embedded stage when clusters in their infancy are surrounded by large quantities of gas and dust \citep[as has been recently demonstrated with JWST NIRCam observations of several nearby galaxies:][]{rodriguez_phangs-jwst_2022, whitmore_phangs-jwst_2023}. Acquiring a detailed understanding of this early phase of cluster evolution can offer valuable insights into the importance of massive star feedback \citep[e.g,][]{bik_sequential_2010}{}{}, as well as its potential role in halting or triggering further star formation surrounding the central source within a giant molecular cloud \citep[GMC;][]{bcm11}{}{}.

Determining the clearing timescale of the gas in individual galaxies provides a direct constraint for the mechanisms predominately responsible for the transition from embedded clusters to partially-embedded and exposed clusters in these systems \citep[][]{reines08, whitmore_using_2011, hollyhead_studying_2015, grasha_connecting_2018, hannon_h_2019, hannon22}{}{}. The most massive young, compact, and embedded star clusters (YSCs) are often considered to be the precursors of the globular clusters observed today \citep[e.g,][]{schweizer87,schweizer98,bcm99,johnson15,kruijssen_globular_2015}{}{}. Thus, understanding the evolution of YSCs in the local Universe can shed light on star cluster formation and evolution across cosmic time.

The target for this study is the “Southern Pinwheel” galaxy, M83 (NGC 5236). It is the nearest, face-on grand-design spiral galaxy at a distance of 4.5 Mpc \citep[][]{thim_cepheid_2003}{}{}. It exhibits a slightly barred structure and has a rich and varied cluster population that has been observed in multiple studies \citep[e.g,][]{rc10c, nb12}{}{}. Moreover, outside of its central starburst, M83 bears significant resemblance to our own Milky Way galaxy, allowing us to systematically study cluster formation and evolution across the disk of a normal star-forming galaxy.

Many previous studies \citep[e.g,][]{rc10c, nb12, fouesneau12, esv14, rc14, ryon15, aa15b} {}{} have focused on clusters inside M83 using broadband wavelengths. These papers examined the Ultraviolet (UV) to Near Infrared (NIR) spectral energy distributions (SEDs) of clusters (using observations from ACS and WFC3/UVIS), and were able to analyze star cluster populations and the distributions of cluster mass, age, luminosity, and size. \cite{fouesneau12} was able to develop a stochastic model to derive age-mass distributions for very young and low mass clusters. Several studies have examined the cluster age distributions finding evidence both for \citep{esv14} and against \citep{rc10c} environmentally-dependent cluster disruption in M83.

Few studies, such as \cite{liu_extinction_2013}, have explored narrowband wavelengths - using data from HST observations, yet without focusing on stellar clusters. For example, \cite{liu_extinction_2013} explores the Hydrogen Alpha (H$\alpha$) and Pa$\beta$ recombination line emission to investigate the surrounding dust geometry and distribution of star-forming regions and stellar complexes; and was able to improve the dust extinction corrections for spatially-resolved pixel-based analyses of galaxies similar to M83.

In this study, we combine the broadband UV-NIR SEDs from WFC3/UVIS along with the narrowband Pa$\beta$ observations from WFC3/IR to identify and characterize the youngest and most NIR-bright star clusters in M83. By doing so, we will robustly sample the earliest phase of cluster formation, where sources have been sparsely seen with existing datasets focusing on UV/optical emission alone \citep[e.g. the lower panel of figure 10 in][]{rc10c}{}{}, filling out the gap in the plots at very young ages. By focusing on this young age range, we can further understand the early stages of cluster evolution and extend the trend lines found by the older literature.

The paper is structured as follows: Section 2 provides information about our observations. Section 3 presents the results obtained from our study. Section 4 engages in a discussion of the obtained results, and Section 5 offers a summary of the primary findings of the research.

\begin{figure*}
\centerline{\includegraphics[width=7truein]{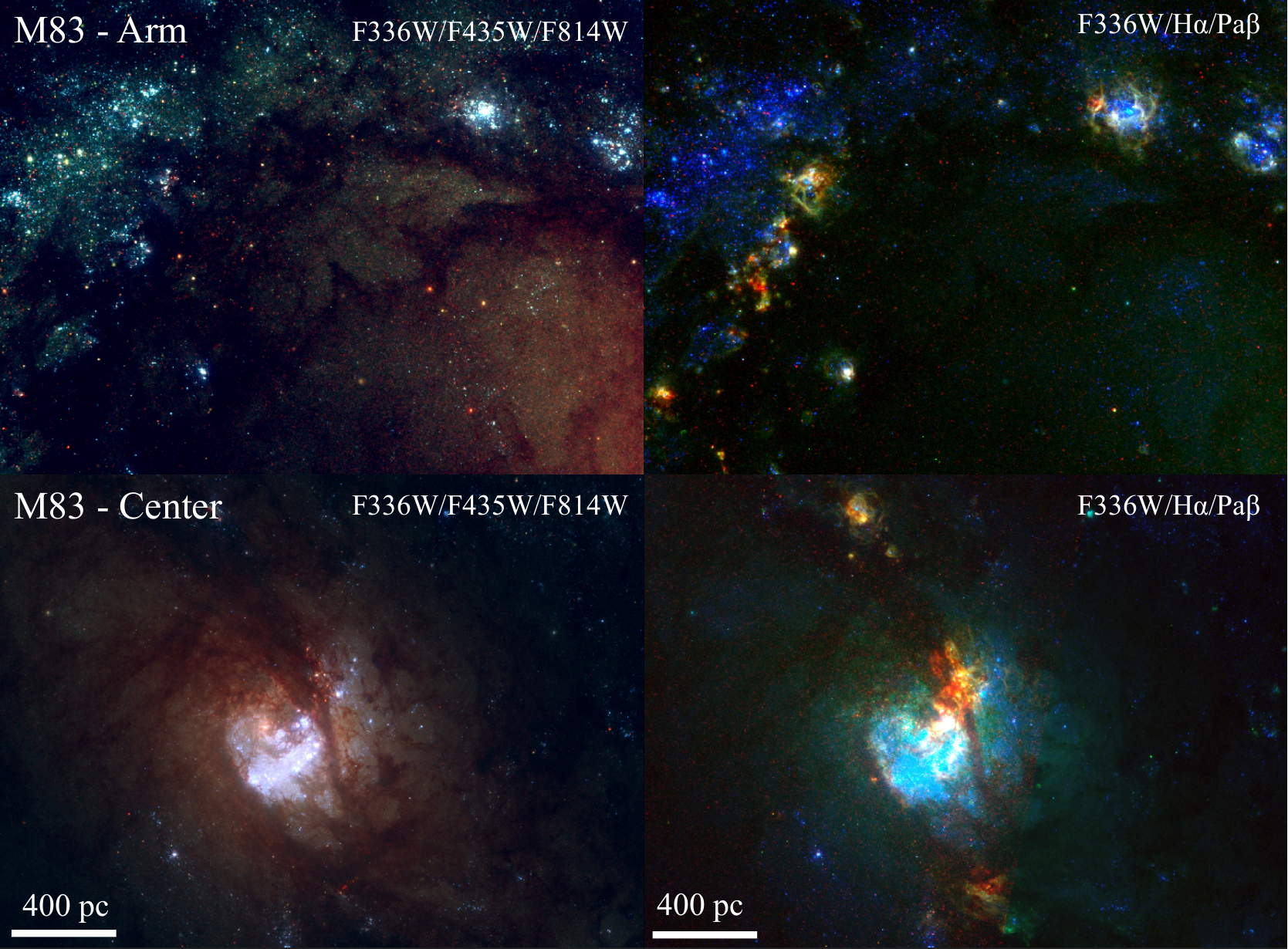}}
\caption{\label{false-color}~False-color RGB (F814W/F435W/F336W) and F446W/H${\alpha}$/Pa${\beta}$ images of the central region ({\bf Bottom}) and inner spiral arm ({\bf Top}) in M83. The left Panels show UV-optical emission which highlight the relatively bright, dust-free, and intermediate-age stellar populations in the disk. The right Panels focus on optical and near-IR Hydrogen recombination line emission (H${\alpha}$ and Pa${\beta}$) which highlight the youngest and most heavily dust-enshrouded stellar clusters. The combination of both broad- and narrow-band imaging allow us to robustly identify and characterize these very young sources across M83 for the first time.}
\end{figure*}

\section{Observations}

The data utilized in this paper were obtained from observations in the Early Release Science project 1 (PI: O'Connell - Program 11360), taken in August 2009. Using the Wide-Field Camera 3 (WFC3) with the Ultraviolet Imaging Spectrometer (UVIS), M83 was observed with the F336W, F438W, F547M, and F657N filters at 0.039''/pix. The WFC3's Infrared (IR) mode provided observations in the F110W, F128N, and F160W filters at 0.128''/pix, and the Advanced Camera for Surveys (ACS) Wide Field Channel (WFC) observed M83 in the F555W and F814W filters at 0.05''/pix. In total, we utilize 9 filters spanning the NUV-NIR, including two hydrogen recombination lines. For each filter, the standard calibration pipeline CALWFC3 v3.1.6 was used to process individual frames into the final images, after correction for bias, dark, and flat fields. All images were aligned to the Gaia DR2 reference frame, and then re-sampled to 0.04''/pix and 0.08''/pix for ACS/UVIS and IR observations respectively \citep{gaiadr2}.

We produce emission line images by subtracting the stellar continuum from the narrow-band F658N and F128N filters. For the F658N image we use the F555W and F814W filters to perform a two point interpolation of the continuum flux at the central wavelength of F658N. For the F128N image we use the F110W and F160W images to perform the interpolation of the continuum flux at the central wavelengths of F128N. Both of these continuum images are then scaled using isolated foreground stars within the F658N and F128N images which do not have any contribution from nebular emission. We further remove the contribution of $N[II]$ to the F658N continuum-subtracted flux using the average $N[II]/H\alpha$ ratio adopted for M83 of 0.53 \citep{kennicutt08,bresolin16}. The presence of Pa$\beta$ in the F110W filter requires the subtraction to be performed in two iterations by subtracting the line emission map from the original F110W observation to create a line-free 1.1 $\mu$m continuum image to be used in the interpolation of the continuum flux at the central wavelength of the F128N image.

For a consistency check at the end, we employed a stellar-continuum subtracted Pa$\alpha$ ($\lambda$ 1.8756~$\mu$m) line emission map acquired using the JWST NIRCam F187N filter, and adjacent broad-band filters (F150W and F200W - priv. comm., A. Adamo 2024) to confirm that our sources are genuine candidates.

In Figure \ref{false-color} we show false-color zoom-ins into different parts of M83 including a dusty-spiral arm and the galaxy center. As we move from UV/optical images to observations of the recombination line emission we see a progression of HII regions emerging along the spiral arm. From bottom left to top-right the HII region bubbles begin as red compact objects (bright in Pa$\beta$ emission), which appear hidden in the UV/optical images, and as the clusters emerge they become progressively larger and dominated by $H\alpha$ emission in their outer shells. This sequence highlights the need for both NUV-NIR continuum as well as deep hydrogen recombination line maps to properly track star cluster formation and evolution in the disks of nearby galaxies.

\begin{figure*}
\centerline{\includegraphics[width=5truein]{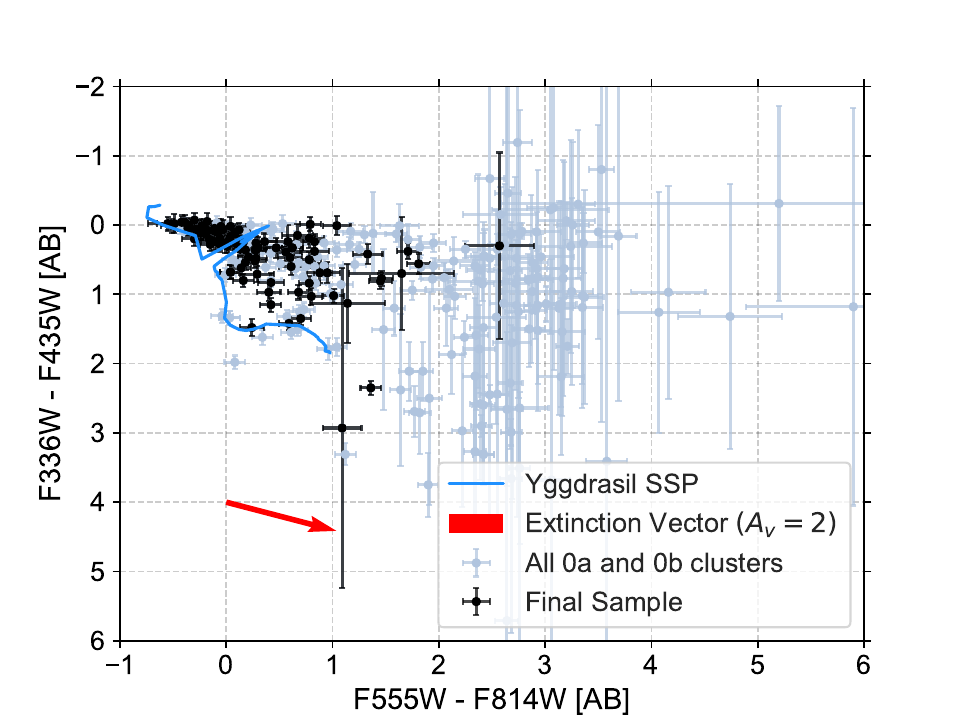}}
\caption{\label{phot_graph} The F336W - F435W vs. F555W - F814W color-color diagram for all visually classified Class 0a and 0b clusters (N = 270 - light grey). Overlaid in light blue is the Yggdrasil SSP models adopted here ($f_{cov} = 0.5$, Kroupa IMF, $z=0.02$) to fit cluster ages, masses, and extinctions. The cluster catalog after applying all selection cuts described in Section 2.2 are show in black (97 sources). In red we show the magnitude of an $A_{V} = 2$ extinction vector, demonstrating that our selection method can identify sources with F555W - F814W up to $\sim 1.5$}, which are likely heavily dust-enshrouded and missed in previous UV/optical surveys of the star cluster populations in M83.
\end{figure*}

\subsection{Cluster Identification, Photometry, and Model-Fitting}

Within the area of M83 determined by the field of view of the two NIR pointings, we identify sources that have compact Pa$\beta$ recombination line emission and, therefore, are possibly YSCs embedded in dust. YSC candidates are extracted from both the ACS F814W and the continuum-subtracted Pa$\beta$ line emission image at their respective native pixel scales. The following procedure is modeled after the Legacy ExtraGalactic UV Survey \citep[LEGUS - ][]{calzetti15,aa17} {}{}

We use SExtractor (Bertin \& Arnouts 1996) with input parameters to identify sources with a S/N $\geq 10$ relative to the local background in 27 contiguous pixels with 50 deblending sub-thresholds, and a background mesh of 100 pixels in the F814W image. For the Pa$\beta$ image we identify sources with a S/N $\geq 5$ relative to the local background in 12 contiguous pixels with 50 deblending sub-thresholds, and a background mesh of 100 pixels. These requirements yield 14991 and 2130 detections in the F814W and  Pa$\beta$ images respectively. This procedure returns the positions of candidate clusters within each pointing and the concentration index (CI: defined here as the magnitude difference of the source within apertures of 1 and 3 pixels and 0.5 and 1.5 pixels for the WFC3/IR images). The CI is related to the effective radius of each object \citep{ryon17} and can be used to differentiate between individual stars and clusters (we specifically selected for CI between the ranges of 1 and 3). We then match the two catalogs within a 1.5 pixel tolerance (1.3 pc), retaining 482 sources in common. Removing obvious edge artifacts, and stars which are over/under-subtracted (producing negative bowls around sources) in the Pa$\beta$ image reduces the total number of sources to be visually-classified to 454.

Standard aperture photometry is performed for each YSC candidate using a fixed 4 pixel radius with a local sky annulus from 5-7 pixels in all 6 NUV-optical filters, and a 2 pixel radius with an annulus from 2.5-3.5 pixels in the 3 IR filters. Aperture corrections to account for missing flux are based on isolated clusters in each pointing and calculated by subtracting the photometry in our fixed apertures from the total photometry inside a 20 pixel and 10 pixel radius aperture with an adjacent 1 pixel-wide sky annulus for the NUV-optical and IR filters respectively. Finally, corrections for foreground Galactic extinction \citep{schlafly11} are applied to the photometric measurements.

Finally, the equivalent width of each YSC candidate is measured as described in \citep{mm21}, by taking the aperture photometry for each adjacent filter (F555W and F814W for $H\alpha$ and F110W and F160W for Pa$\beta$), interpolating to the central wavelength of the narrow-band filters, and subtracting this estimate for the stellar continuum from the measured flux in the F658N and F128N respectively. The equivalent width is then measured as the ratio of the line and continuum flux values multiplied by the filter bandwidth. 

The SED fitting is performed using the Yggdrasil single stellar population (SSP) models, which implements Cloudy \citep{cloudy23} to include the contribution from nebular emission lines \citep{yggdrasil}. We adopt a \citet{pk01} initial mass function (IMF), Padova isochrones that include thermally pulsating asymptotic giant branch stars \citep{vazquez05}, as well as a MW attenuation curve \citep{cardelli89}. Further, we adopt solar metallicity ($z = 0.02$) models for M83 which are valid for star-forming regions beyond $\sim 0.2 R/R_{25}$ \citep{hernandez19}. Finally we allow for a maximum color excess of E (B-V) = 2.5 to generate our model grid and fit cluster age, mass, and extinction. To perform least-$\chi^{2}$ SED fitting we include significant detections from all 9 filters from the NUV to the NIR, which produces average uncertainties of 0.2 dex for all derived parameters. Since our cluster photometry is measured by applying a single average aperture correction in each filter, its uncertainty affects only the normalization of the SED, and not the overall shape.

\subsection{Visual Classification and Final Catalog Selection}

For all 454 YSC candidates identified the sources are manually classified based on the following criteria:

\vspace{0.05in}

\noindent \textbf{Class 0}: Sources which appear isolated within 4 pixels in both the F814W and continuum-subtracted Pa$\beta$ images are assigned a Class of 0. When further examining the SEDs of these Class 0 sources, we further separate candidates into 0a: Sources with SEDs which appear dominated by a single stellar population, and 0b: sources with SEDs which appear to include one or more stars of different age or color. These very close star/cluster pairs are hard to identify morphologically at the resolution of the available {\it HST} imaging, but appear as clear superpositions of two or more stellar populations, as these sources will not have smoothly-rising or falling SEDs from the near-UV to the near-IR.

\noindent \textbf{Class 1}: Sources which appear to have one or more companions within 4 pixels, which causes contamination in the resulting SED, and results in significant errors on the derived cluster properties. Although a full accounting of the star and cluster formation in M83 includes these close pairs, we exclude them from the remainder of our analysis due to the uncertainties in evaluating the physical properties of multiple blended sources.

\noindent \textbf{Class 2}: Sources with emission which does not appear cluster-like. These candidates are detected as diffuse recombination-line emission regions with a nearby, un-associated stellar source detected at F814W. Further, these sources may be the result of poor subtractions which were not removed in the previous round of quality checks, edge detections, diffraction spikes, or un-removed cosmic rays. Like Class 1 sources we exclude these sources from the remainder of our analysis as they are likely not genuine YSC candidates.

\vspace{0.05in}

\begin{figure*}
\centerline{\includegraphics[width=6truein]{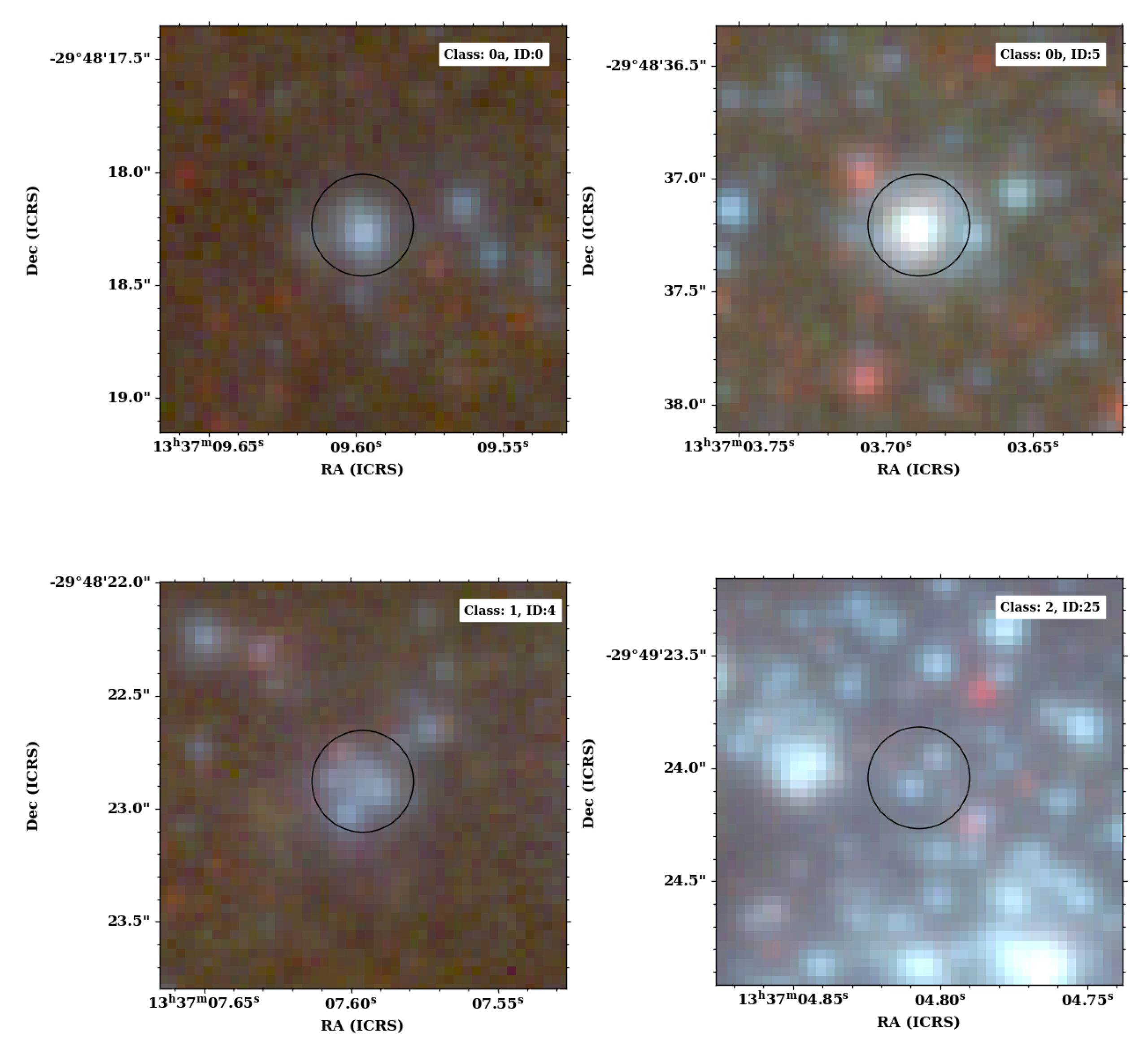}}
\caption{\label{morphology-examples} Three-color images of representative YSC candidates for each of our 4 defined classes: 0a, 0b, 1 and 2. All 454 sources were manually classified based on the criteria mentioned in Section 2.2.}
\end{figure*}

Table 1 lists the number of sources assigned to each classification, as well as their median SED-derived age, mass, and E(B-V) values. Figure \ref{morphology-examples} showcases the morphology differences between all the classes. Following the manual classification, we select our final sample of YSC candidates from Class 0a or 0b sources and remove any which have large NUV photometric uncertainties ($\sigma_{F336W} \geq 0.2$ mag). The F336W filter provides a much improved lever arm for age-dating stellar clusters relative to optical filters alone, and therefore our cut ensures that the reddening-age degeneracies can be reliably broken. We further require H$\alpha$ and Pa$\beta$ equivalent widths greater than 0, and sources with non-detections in less than 2 bands. These cuts remove an additional 165 sources, retaining a sample of 105 YSC candidates. Although these cuts likely exclude some heavily dust-enshrouded sources in the WFC3-IR footprint, these clusters have very uncertain NUV photometry (and in some cases Pa$\beta$ photometry as well), and thus it is hard to derive ages/extinction values from SED fitting \citep[e.g. see][]{aa17,linden22}.

Finally, we compare the locations of all 105 YSC candidates to a stellar-continuum subtracted Pa$\alpha$ ($\lambda$ 1.8756~$\mu$m) line emission map acquired using the JWST NIRCam F187N filter, and adjacent broad-band filters (F150W and F200W - priv. comm., A. Adamo 2024). This image has 2x better angular resolution (0.04''/pix) relative to the HST/WFC3/IR observations used here, and thus provides a final consistency check that the sources extracted with Pa$\beta$ line emission are genuine YSC candidates. By visually examining the Pa$\alpha$ emission at our YSC locations we find that 8/105 candidates appear as subtracted stars at the resolution of JWST NIRCam. We therefore remove these infrared-selected sources from our sample catalog, leaving us with a final sample of 97 YSC candidates.

For the remainder of the paper we focus on this sub-sample which represents our best catalog of infrared-selected star clusters in M83 with {\it HST}. Such rigorous cuts are implemented in order to provide a clean sample to examine how young star clusters evolve over their first few Myr. The positions, classifications, and SED-derived physical properties for all 97 YSC candidates are given in Table 2.

\begin{figure*}
\centerline{\includegraphics[width=9truein]{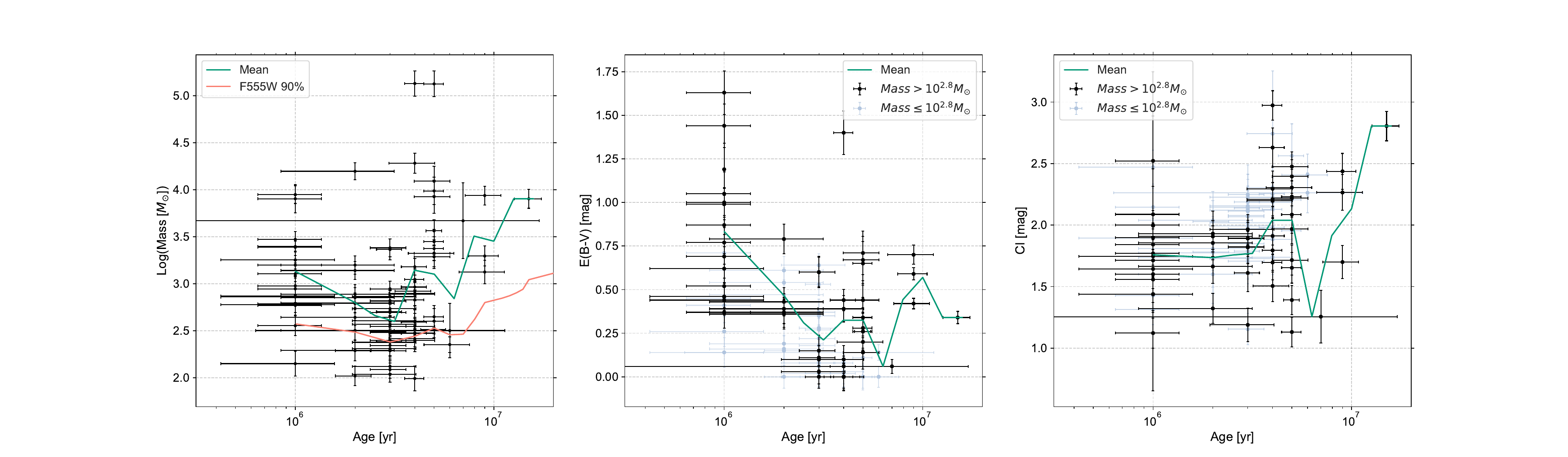}}
\caption{\label{multi-graph}{\bf Left}: The age and mass for our final sample of 97 Class 0a and 0b clusters. The solid-red line represents the F555W 90\% completeness limit derived using the LEGUS Cluster Completeness Tool \citep[$M_{V} = -6.24$;][]{linden22} {}{}. The goal of this study is to evaluate the population of clusters in M83 with ages less than 10 Myr. Therefore to achieve a mass-limited sample over this entire age range \citep[see][]{renaud20} we impose a cut of $M > 10^{2.8} M_{\odot}$, which removes an additional 44 sources from the final analysis. {\bf Middle}: The Extinction vs. age for our 53 most massive clusters in M83. The lower-mass sources removed to maintain a mass-complete sample over the first 10 Myr are shown in grey. The solid-green curve represents the running mean of the distribution. Our Pa$\beta$ selection has increased the number of young ($t \leq 3$ Myr) and dusty ($E(B-V) > 0.5$) sources by 14 relative to previous studies of M83 \citep[][]{nb12, rc14}{}{}. {\bf Right}: The CI vs. age for our 53 most massive clusters in M83. The lower-mass sources removed to maintain a mass-complete sample over the first 10 Myr are again shown in grey. The solid-green curve represents the running mean of the distribution. We do not see a significant size evolution over the age-range studied here. This result is broadly consistent with \citep[][]{ryon15}{}{} who show modest to little evolution of cluster size in the first 10 Myr.}
\end{figure*}

\begin{table}
\begin{ruledtabular}
\begin{tabular}{cc}
 \multicolumn{2}{c}{\textbf{Class 0a}}\\
 \# of clusters identified&172\\
 Median Age [log(Yr)]&6.778\\
 Median Mass [log(\(M_\odot\))]&3.588\\
 Median E (B-V) [Mag]&0.460\\ \hline
 \multicolumn{2}{c}{\textbf{Class 0b}}\\
 \# of clusters identified&98\\
 Median Age [log(Yr)]&6.778\\
 Median Mass [log(\(M_\odot\))]&3.742\\
 Median E (B-V) [Mag]&0.895\\ \hline
 \multicolumn{2}{c}{\textbf{Class 1}}\\
 \# of clusters identified&136\\
 Median Age [log(Yr)]&6.699\\
 Median Mass [log(\(M_\odot\))]&3.423\\
 Median E (B-V) [Mag]&0.510\\ \hline
 \multicolumn{2}{c}{\textbf{Class 2}}\\
 \# of clusters identified&48\\
 Median Age [log(Yr)]&6.778\\
 Median Mass [log(\(M_\odot\))]&3.986\\
 Median E (B-V) [Mag]&1.270\\ \hline
 \multicolumn{2}{c}{\textbf{Final cluster catalog}}\\
 \# of clusters identified&97\\
 Median Age [log(Yr)]&6.477\\
 Median Mass [log(\(M_\odot\))]&2.858\\
 Median E (B-V) [Mag]&0.260\\
\end{tabular}
\end{ruledtabular}
\caption{\label{tab:cluster_stats} Population statistics for the 454 sources with S/N $\geq 3$ in F814W and Pa$\beta$ imaging, as well as convergent UV-NIR SED fits. The last section gives population statistics for the final 97 clusters left after applying further classification cuts.}
\end{table}
 
\section{Results}

We have presented an effective method for identifying and classifying star clusters in M83, allowing us to specifically target infrared-bright YSCs. In Figure \ref{phot_graph} we present the F336W-F435W vs F555W-F814W color-color diagram for all 270 isolated and compact sources (Classes 0a and 0b) as well as our final sample of 97 confirmed YSC candidates. From the Yggdrasil SSP model shown we find a tight population of young and blue sources with F336W-F435W $\sim 0.8$ and F555W-F814W $\sim 0$ as well as a redder population with F555W - F814W $\geq 0.5$. For reference we also plot an extinction vector with A$_{V} = 2$ demonstrating that these very red sources have a color excess E(B-V) $\sim 1$. These sources have largely been overlooked in previous UV/optical surveys.

In Figure \ref{multi-graph} we show the age vs. mass, age vs. E(B-V) and age vs. Concentration Index (CI) for the 97/270 clusters which pass all selection criteria. In the left Panel of Figure \ref{multi-graph} we show the F555W 90\% completeness limit for our observations determined through injection and recovery of synthetic sources across our data frames (see \citep{aa17} and \citep{linden22} for details on the LEGUS Cluster Completeness tool). At an age of 10 Myr, we observe a 90\% completeness limit in F555W at a mass of $10^{2.8}$ solar masses (\(M_\odot\)). A final catalog cut removed all sources less than $10^{2.8}$ \(M_\odot\) and left us with 53 of the most massive and youngest clusters. The black points on the central and right panels of Figure \ref{multi-graph} are these 53 clusters with masses greater than $10^{2.8}$ \(M_\odot\), while the lighter gray points are the 44 clusters below that cutoff. We only used the 53 final data points to plot the mean curve on the age vs. E(B-V) and age vs. CI graphs.

In the middle Panel of Figure \ref{multi-graph} we show the color excess E(B-V) vs cluster age. Over the first $\sim3$ Myr we can clearly see the extinction decrease as the age of a cluster increases. \cite{reines08, whitmore_using_2011, hollyhead_studying_2015, hannon_h_2019, hannon22} concluded that YSCs start removing gas at around 2-3 Myr and become free of gas at $\sim4$ Myr. We can measure what percentage of clusters have an $A_V$ less than 1 as age increases. We find that by 2 Myr 75\% of clusters have an $A_{V} < 1$, and that by 3 Myr the percentage rises to 82\%. These results match the literature cited above, with clusters removing gas starting before 2 Myr and a large majority becoming relatively dust free by 3 Myr. Thus, we can provide a strong constraint: the clearing timescale for YSCs (that are detected in the UV/optical) in M83 is $\lesssim 3$ Myr. 

There is, however, a single high-mass ($\sim 10^{5} M_{\odot}$) cluster with high extinction ($A_{V} \geq 1.4$ mag) with an age of 4 Myr. This source seems to be the exception as the majority of the clusters lose their dust by this time, but the existence of dusty YSCs older than 4 Myr is at odds with existing models that work with pre-SN feedback \citep{krumholz_star_2019}. This is similar to cluster candidates found in \cite{calzetti_dust_2023}, and would imply that supernova feedback is required to fully clear the gas and dust around these YSCs. However, with only a single source in this regime we are unable to determine whether there is a strong mass-dependence to the clearing timescale for clusters in M83.

\begin{figure*}
\centerline{\includegraphics[width=8truein]{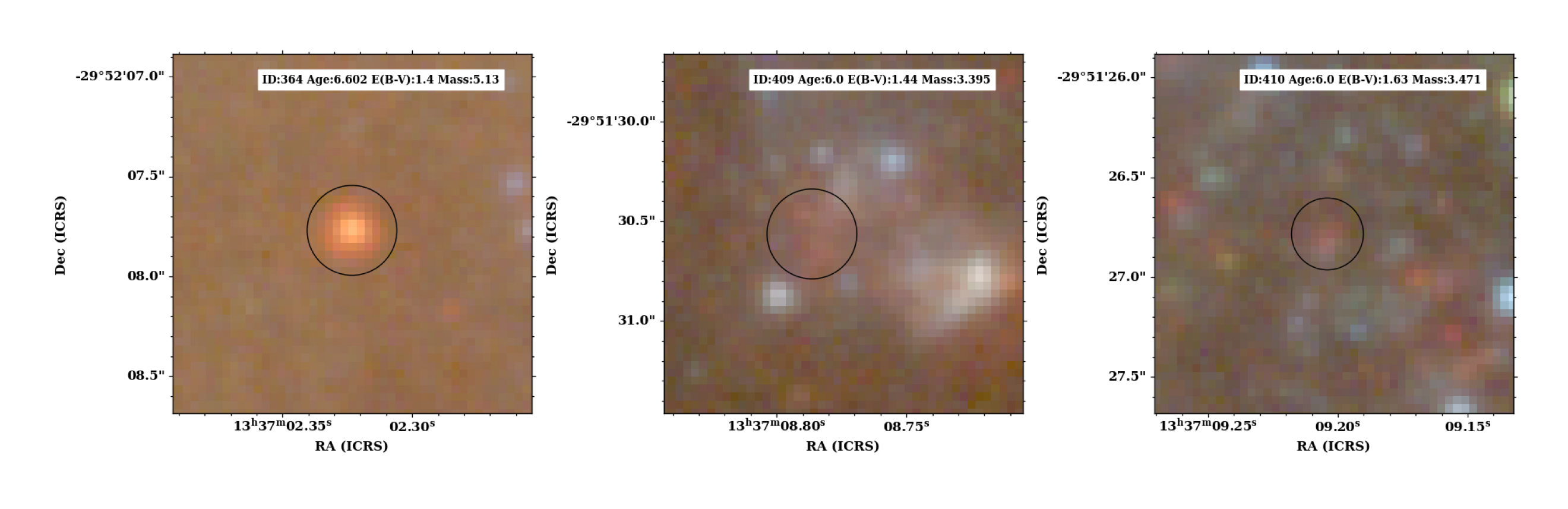}}
\caption{\label{oldest} The three YSCs in our final cluster catalog with the highest SED-derived E(B-V) values. The leftmost Panel shows an older source - ID 364 - who has age $\geq 4$ Myr, $M_{*} \geq 10^{4.7} M_{\odot}$, and E(B-V) $\geq 1.4$. Although rare within our WFC3 footprint, this source may represent an important population of clusters requiring supernova feedback to fully disperse their surrounding gas and dust across the full disk of M83. The middle and right Panels show the younger two sources with ages $\sim 1$ Myr, $M_{*} \sim 10^{3.4} M_{\odot}$, and E(B-V) $\geq 1.4$. These very young and heavily extincted sources are the signposts for active feedback into the surrounding ISM.}
\end{figure*}

In Figure \ref{oldest} we show false-color F435W/F555W/F814W cut-outs for the 3 YSC candidates with the highest E(B-V) values identified in our catalog. The oldest source with an age of $\sim 6$ Myr (left Panel) is clearly a red, compact, YSC with detections of both emission lines, and a $\sigma_{F336W} < 0.2$. Further, we find that this source resides within the nuclear region of the galaxy where intense starburst activity is seen \citep{gallais91}. We stress that the population of these sources is very small within the footprint of M83 studied here, however across the entire galaxy there may be a larger population of YSCs where supernova feedback will be required to fully clear the surrounding gas and dust.

In order to determine the impact our selection method has on the resulting YSCs detected with both compact Pa$\beta$ emission and $\sigma_{F336W} < 0.2$ we examine the distribution of U-band photometric magnitudes vs. uncertainties, and find that sources with $m_{U} > 24$ have errors larger than 0.2 dex. Further, we find that the brightest sources in our catalog reach $m_{U} \sim 19$. If we assume that these sources are dust-free, our photometric selection ($\sigma_{F336W} < 0.2$) is sensitive enough to detected clusters in M83 with a maximum E(B-V) of 1.61, and an average E(B-V) of $\sim$ 1 for clusters with $M > 10^{2.8} M_{\odot}$. These limits are sufficient to properly sample IR-bright star clusters detected across the disk of M83.

Finally, the right Panel of Figure \ref{multi-graph} shows the concentration index vs cluster age. Changes in CI match previous observations in the LEGUS survey \cite{brown_radii_2021}, which found a clear positive correlation for clusters older than 10 Myr, but a very shallow and weakly positive correlation for clusters younger than 10 Myr. This was also seen in \cite{ryon15} for optically-selected sources in M83. Our data is consistent with these findings given that the running median for infrared-selected sources stays constant over the first $\sim 5$ Myr. This indicates that there is no significant early evolution in size for clusters with masses $>10^{2.8}$ \(M_\odot\) in M83.

\section{Discussion}

\subsection{Spectroscopic Comparison}

In order to independently verify the results of our SED fitting we cross-references our final YSC catalog with spectroscopic samples of star clusters from previous work in the literature. From \cite{wofford_ultraviolet_2011} we found 2 common sources: cluster \#82 in our catalog corresponds to cluster \#1 in their paper, and cluster \#399 is cluster \#14. For the first cluster we derived age of 4.0 Myr and E(B-V) of 0, which is consistent within uncertainties with the age and extinction derived from UV spectroscopy (3.9 Myr and E(B-V) = 0.1). For the second cluster we derive a SED-fitted age of 3.0 Myr and an E(B-V) = 0, which again is consistent within the measured uncertainties to the values derived from UV Spectroscopy (2.2 Myr and E(B-V) = 0.2). This comparison demonstrates that for sources which have been targeted for spectroscopic follow-up, our SED fitting technique does a very good job at accurately recovering their physical properties. This provides further evidence that the cloud of sources we identify with E(B-V) = 0 with ages of $\sim 3$ Myr is a real population in M83.

\subsection{YSC Clearing Timescales}

Various studies of star clusters and GMCs in nearby galaxies support different conclusions about the role of supernova (SN) or pre-supernova (pre-SN) feedback as the dominant mechanism for YSCs to clear their surroundings. \citet{grasha19} used the associations of YSCs and molecular gas to derive a timescale of $\sim$ 6 Myr for star clusters to become exposed in the disk of M51. \citet{matthews_resolved_2018} used HST and ALMA CO (3-2) observations of the Antennae merger to demonstrate that some clusters begin to remove their molecular material at ages $\leq$ 5 Myr, and that by 30 Myr virtually all sources appear to be gas-free. More recently, \citet{mcquaid24} used UV/optical observations with HST to derive a clearing timescale of 5-6 Myr for infrared-selected star clusters in NGC 4449. They find hints that the most massive line-emitting sources in the galaxy clear their surroundings faster ($\sim 4$ Myr) than their lower-mass counterparts, and that YSC clearing timescales may be mass-dependent.

However, several other works find timescales of $<$ 4-5 Myr for recombination line-emitting star clusters in nearby disk galaxies, supporting the idea that pre-SN feedback, as the earliest supernovae occur at around $\sim4$ Myr - too late to have caused the clearing of dust from majority of the YSCs, to be a major factor in regulating star formation \citep{whitmore_using_2011, hollyhead_studying_2015, sokal15, sokal16, hannon_h_2019, hannon22}. Importantly however, many of these studies are based on UV/optical data alone. Crucially, \cite{messa_looking_2021} demonstrated that standard UV/optical catalogs miss about 40\%-60\% of star clusters younger than 6 Myr in NGC 1313, while \cite{linden_goals-jwst_2023} (which uses JWST NIRCam F150W, F200W, F356W, and F444W observations) inferred that up to 2/3 of clusters younger than 4 Myr are missed at UV wavelengths in more luminous starburst galaxies in the local Universe.

Overall, we have 76 sources that are younger than 5 Myr, out of which 58 are younger than 3 Myr. When compared to previously published catalogs of star clusters in M83 we have increased the total number of YSCs detected in the age range of 1 - 3 Myr by 3x relative to Figure 5 in \citet[][]{nb12}. This is due in large part to the use of the Pa$\beta$ emission line map, which was not used in any previous study of star clusters in M83, and demonstrates that NIR recombination lines are crucial for recovering the youngest and dustiest sources in nearby galaxies. Overall, our Pa$\beta$ selection has increased the number of young ($t \leq 3$ Myr) and dusty ($E(B-V) > 0.5$) sources by 14 relative to previous studies of M83 \citep[][]{nb12, rc14}{}{}

Our inferred clearing time of $\sim$ 3 Myr implies that pre-SN feedback plays a key factor in star feedback and regulation of star formation. As noted in \cite{krumholz_star_2019}, the greatest component of the pre-SN feedback for YSCs with mass $< 10^{4} M_{\odot}$ is direct radiation pressure. Therefore, it is possible that direct radiation pressure may be a dominant mechanism for stellar clusters with M $< 10^{4} M_{\odot}$ in the disk and central regions of M83. We also note that we do not see any evidence for a mass-dependence in the clearing timescale for infrared-selected sources in M83.

A point to consider here is the effect of metallicity on the pre-supernova feedback, as metallicity positively correlates with the strength of stellar winds and hence has an effect on the duration of the clearing timescale \citep{jecmen23}. In this case, the radial gas-phase metallicity rises steeply in the central region of M83 to super-solar values, and smoothly declines in the outer-disk beyond 2-3 kpc \citep{hernandez19}. When we separate out YSC candidates into the inner- and outer-disk pointings we find that YSCs in both regions occupy approximately the same distributions of E(B-V) and age, suggesting that the effects of combining YSCs from regions of M83 with different metallicity does not strongly bias the results. Further exploring what feedback mechanisms may dominate at different galactocentric radii will be addressed with future JWST NIRCam observations, which span nearly the entire disk due to the large FOV, and allows us to probe a much larger range of metallicities and gas densities relative to the individual HST WFC3-IR pointings.

Finally, to further determine what early feedback mechanisms are important for our infrared-selected sources in M83, we follow the calculations presented in \cite{lopez14} to determine the radiation pressure in the HII regions surrounding these YSCs. From our derived physical properties of our YSC candidates, we adopt an average E(B-V) of 0.5, and a median H$\alpha$ flux of $5$x$10^{-18}$erg/s/cm$^{2}$/$\AA$. We further adopt a median cluster radius of 3pc \citep{ryon15}, to derive a value for the direct radiation pressure $P_{dir}$ of $3.8$x$10^{-10}$ g/cm/s$^{-2}$. This value is comparable to typical ionization ($P_{ion}$) pressures found for HII regions in the LMC of $\sim 6$x$10^{-10}$g/cm/s$^{-2}$ \citep{lopez14}. This is consistent with the results presented in \citet{della-bruna22}, which used MUSE to investigate the direct and ionization pressures for optically-selected HII regions in the disk and central regions of M83. They find that for the dustiest HII regions $P_{dir}$ is approximately constant with galctocentric radius, and becomes comparable in strength to $P_{ion}$ in the central region. Therefore we believe that direct radiation pressure must play an equally important role in the early feedback process for YSCs in the disk of M83.

\section{Summary}
We present an analysis of Hubble Space Telescope data which uses both broadband and narrow-band wavelengths to study the young star cluster population of the face-on spiral galaxy M83. Starting with 454 sources identified with continuum subtracted H$\alpha$ \& Pa$\beta$, we manually classified clusters based on their spectral energy distributions and multi-wavelength morphology. From the 270 sources classified as either 0a or 0b, we further remove 173 sources with H$\alpha$ and Pa$\beta$ equivalent widths $< 0$, sources with $\sigma_{F336W} > 0.2$, any sources with upper-limits measured in more than one band, and sources which do not have a Pa$\alpha$ counterpart in existing JWST NIRCam imaging, leaving us with 97 YSC candidates. A 90\% completeness curve in F555W was applied which resulted in 53 clusters with a mass $> 10^{2.8} M_{\odot}$. Based on analysis of this final catalog, we reached the following conclusions:

(1) We find by 2 Myr 75\% of our infrared-selected sources have an $A_{V} < 1$, and that by 3 Myr the total reaches 82\%. Our result matches previous studies \citep{hollyhead_studying_2015, hannon_h_2019, hannon22}, and provides a stronger constraint on the clearing timescale for YSCs in M83: Clusters start removing gas before 2 Myr, and a large majority become relatively dust free by 3 Myr. This implies that pre-supernovae feedback is the primary mechanism for dust clearing with only a small group clusters remaining IR-bright long enough for supernovae to play an important role.

(2) We find a weak and shallow direct relation between size (CI) and age in the first 10 Myr. It matches studies done on the LEGUS survey \citep{brown_radii_2021} and indicates there is not much change in size during this time frame. However we stress that our infrared-selected sources may still exclude some of the most embedded clusters, since our selection criteria require our sample to be well detected at F336W and longer bands. 

\begin{acknowledgements}

The authors thank the FEAST team (JWST Cycle 1 program 1783) for providing the stellar–continuum subtracted Pa$\alpha$ image of M83. The results of this paper are based on observations made with the NASA/ESA Hubble Space Telescope, obtained at the Space Telescope Science Institute, which is operated by the Association of Universities for Research in Astronomy, Inc., under NASA contract NAS 5–26555. The data presented in this paper were obtained from the Mikulski Archive for Space Telescopes (MAST) at the Space Telescope Science Institute. The specific observations analyzed can be accessed via \dataset[DOI: 10.17909/9e0g-e814]{https://doi.org/10.17909/9e0g-e814}. This research has made use of the NASA/IPAC Extragalactic Database (NED) which is operated by the Jet Propulsion Laboratory, California Institute of Technology, under contract with the National Aeronautics and Space Administration.

\end{acknowledgements}

\bibliography{master_ref}

\begin{longdeluxetable}{|l|ll|l|l|l|l|}
\tablecaption{Final Cluster Catalog \label{tbl-2}}
\tablewidth{0pt}
\tablehead{
\colhead{ID} & \colhead{RA (J2000)} & \colhead{Dec (J2000)}  & \colhead{Class} & \colhead{Age [log(Yr)]} & \colhead{Mass [log(\(M_\odot\))]} & \colhead{Extinction [Mag]}}
\startdata
0 & 204.28999 & -29.80507 & 0a & 6.477 $\pm$ 0.08805 & 2.396 $\pm$ 0.1017 & 0.11 $\pm$ 0.08 \\ 
1 & 204.27233 & -29.80545 & 0a & 6.0 $\pm$ 0.1505 & 3.387 $\pm$ 0.0868 & 0.37 $\pm$ 0.09 \\ 
3 & 204.28584 & -29.80962 & 0a & 6.477 $\pm$ 0.1505 & 2.034 $\pm$ 0.08195 & 0.06 $\pm$ 0.075 \\ 
10 & 204.26184 & -29.81082 & 0a & 6.602 $\pm$ 0.04846 & 2.828 $\pm$ 0.1046 & 0.0 $\pm$ 0.075 \\ 
12 & 204.24918 & -29.81036 & 0a & 6.477 $\pm$ 0.08805 & 2.819 $\pm$ 0.1053 & 0.6 $\pm$ 0.09 \\ 
13 & 204.2492 & -29.8118 & 0a & 6.477 $\pm$ 0.08805 & 2.816 $\pm$ 0.1105 & 0.15 $\pm$ 0.09 \\ 
14 & 204.28468 & -29.8135 & 0a & 6.0 $\pm$ 0.2386 & 2.869 $\pm$ 0.1021 & 0.44 $\pm$ 0.085 \\ 
15 & 204.28248 & -29.81346 & 0a & 6.477 $\pm$ 0.08805 & 2.945 $\pm$ 0.0811 & 0.11 $\pm$ 0.07 \\ 
16 & 204.27638 & -29.81695 & 0a & 6.0 $\pm$ 0.1505 & 2.769 $\pm$ 0.1142 & 0.41 $\pm$ 0.1 \\ 
17 & 204.28537 & -29.81847 & 0a & 6.602 $\pm$ 0.08805 & 2.921 $\pm$ 0.1005 & 0.0 $\pm$ 0.08 \\ 
21 & 204.29153 & -29.81966 & 0a & 6.477 $\pm$ 0.1505 & 2.864 $\pm$ 0.0835 & 0.1 $\pm$ 0.08 \\ 
23 & 204.27268 & -29.82189 & 0a & 6.0 $\pm$ 0.2386 & 3.252 $\pm$ 0.09415 & 0.46 $\pm$ 0.08 \\ 
27 & 204.29146 & -29.82514 & 0a & 6.778 $\pm$ 0.3495 & 2.501 $\pm$ 0.2883 & 0.14 $\pm$ 0.045 \\ 
29 & 204.26404 & -29.82733 & 0a & 6.699 $\pm$ 0.04846 & 3.373 $\pm$ 0.1273 & 0.28 $\pm$ 0.115 \\ 
31 & 204.27844 & -29.82805 & 0a & 6.602 $\pm$ 0.1109 & 2.477 $\pm$ 0.1505 & 0.03 $\pm$ 0.085 \\ 
35 & 204.27881 & -29.82694 & 0a & 6.699 $\pm$ 0.08805 & 3.926 $\pm$ 0.1757 & 0.67 $\pm$ 0.135 \\ 
38 & 204.26239 & -29.82897 & 0a & 6.602 $\pm$ 0.06247 & 3.179 $\pm$ 0.08695 & 0.06 $\pm$ 0.06 \\ 
40 & 204.27347 & -29.83028 & 0a & 6.477 $\pm$ 0.06247 & 2.384 $\pm$ 0.1013 & 0.27 $\pm$ 0.065 \\ 
42 & 204.28691 & -29.83547 & 0a & 6.602 $\pm$ 0.1109 & 2.395 $\pm$ 0.1173 & 0.06 $\pm$ 0.07 \\ 
45 & 204.25123 & -29.83857 & 0a & 6.477 $\pm$ 0.08805 & 2.547 $\pm$ 0.1115 & 0.18 $\pm$ 0.09 \\ 
47 & 204.2793 & -29.83945 & 0a & 6.602 $\pm$ 0.1109 & 2.897 $\pm$ 0.1315 & 0.1 $\pm$ 0.08 \\ 
55 & 204.28647 & -29.84886 & 0a & 6.778 $\pm$ 0.1109 & 2.352 $\pm$ 0.0832 & 0.0 $\pm$ 0.06 \\ 
56 & 204.28629 & -29.84882 & 0a & 6.477 $\pm$ 0.1505 & 2.86 $\pm$ 0.0731 & 0.03 $\pm$ 0.07 \\ 
67 & 204.24631 & -29.85226 & 0a & 6.0 $\pm$ 0.2386 & 2.149 $\pm$ 0.1304 & 0.14 $\pm$ 0.085 \\ 
69 & 204.28234 & -29.85314 & 0a & 6.301 $\pm$ 0.2386 & 2.886 $\pm$ 0.1027 & 0.37 $\pm$ 0.085 \\ 
74 & 204.24682 & -29.8542 & 0a & 6.301 $\pm$ 0.08805 & 2.288 $\pm$ 0.11 & 0.15 $\pm$ 0.085 \\ 
76 & 204.2822 & -29.85403 & 0a & 6.477 $\pm$ 0.1505 & 2.12 $\pm$ 0.07725 & 0.0 $\pm$ 0.055 \\ 
81 & 204.25448 & -29.85885 & 0a & 6.699 $\pm$ 0.08805 & 3.287 $\pm$ 0.1253 & 0.14 $\pm$ 0.095 \\ 
83 & 204.25622 & -29.85895 & 0a & 6.477 $\pm$ 0.08805 & 3.381 $\pm$ 0.0965 & 0.6 $\pm$ 0.085 \\ 
85 & 204.25139 & -29.86352 & 0a & 6.699 $\pm$ 0.04846 & 5.126 $\pm$ 0.1376 & 0.65 $\pm$ 0.125 \\ 
89 & 204.252 & -29.86525 & 0a & 6.602 $\pm$ 0.1109 & 4.282 $\pm$ 0.1067 & 0.0 $\pm$ 0.065 \\ 
91 & 204.28637 & -29.86812 & 0a & 6.301 $\pm$ 0.2386 & 2.502 $\pm$ 0.1061 & 0.61 $\pm$ 0.08 \\ 
94 & 204.28754 & -29.86933 & 0a & 6.301 $\pm$ 0.2386 & 3.198 $\pm$ 0.097 & 0.39 $\pm$ 0.085 \\ 
105 & 204.27239 & -29.80526 & 0a & 6.301 $\pm$ 0.2386 & 2.789 $\pm$ 0.09725 & 0.19 $\pm$ 0.085 \\ 
152 & 204.29153 & -29.81966 & 0a & 6.477 $\pm$ 0.1505 & 2.864 $\pm$ 0.0835 & 0.1 $\pm$ 0.08 \\ 
161 & 204.28397 & -29.81995 & 0a & 6.602 $\pm$ 0.1109 & 2.414 $\pm$ 0.1091 & 0.0 $\pm$ 0.065 \\ 
164 & 204.27268 & -29.82189 & 0a & 6.0 $\pm$ 0.2386 & 3.252 $\pm$ 0.09415 & 0.46 $\pm$ 0.08 \\ 
176 & 204.27037 & -29.82371 & 0a & 6.699 $\pm$ 0.04846 & 3.447 $\pm$ 0.1318 & 0.26 $\pm$ 0.11 \\ 
192 & 204.28038 & -29.82752 & 0a & 6.477 $\pm$ 0.06247 & 2.488 $\pm$ 0.0853 & 0.02 $\pm$ 0.06 \\ 
202 & 204.27909 & -29.82661 & 0a & 6.0 $\pm$ 0.1505 & 3.947 $\pm$ 0.0958 & 1.0 $\pm$ 0.085 \\ 
244 & 204.2803 & -29.84824 & 0a & 6.699 $\pm$ 0.04846 & 2.602 $\pm$ 0.1343 & 0.11 $\pm$ 0.1 \\ 
247 & 204.2851 & -29.84793 & 0a & 6.0 $\pm$ 0.1505 & 3.105 $\pm$ 0.0833 & 0.77 $\pm$ 0.085 \\ 
262 & 204.28841 & -29.84989 & 0a & 6.699 $\pm$ 0.04846 & 2.648 $\pm$ 0.119 & 0.0 $\pm$ 0.075 \\ 
268 & 204.26376 & -29.85048 & 0a & 6.301 $\pm$ 0.08805 & 2.019 $\pm$ 0.1022 & 0.0 $\pm$ 0.065 \\ 
280 & 204.2806 & -29.85169 & 0a & 6.602 $\pm$ 0.1109 & 2.888 $\pm$ 0.1237 & 0.39 $\pm$ 0.075 \\ 
310 & 204.28396 & -29.85514 & 0a & 6.699 $\pm$ 0.08805 & 4.09 $\pm$ 0.1575 & 0.71 $\pm$ 0.125 \\ 
318 & 204.25577 & -29.85714 & 0a & 6.699 $\pm$ 0.08805 & 3.324 $\pm$ 0.1433 & 0.44 $\pm$ 0.11 \\ 
322 & 204.25448 & -29.85885 & 0a & 6.699 $\pm$ 0.08805 & 3.287 $\pm$ 0.1253 & 0.14 $\pm$ 0.095 \\ 
366 & 204.2512 & -29.86596 & 0a & 6.699 $\pm$ 0.04846 & 3.987 $\pm$ 0.1461 & 0.34 $\pm$ 0.115 \\ 
382 & 204.28779 & -29.86949 & 0a & 6.699 $\pm$ 0.1109 & 3.324 $\pm$ 0.07645 & 0.2 $\pm$ 0.06 \\ 
388 & 204.25342 & -29.87639 & 0a & 6.699 $\pm$ 0.03959 & 3.562 $\pm$ 0.1155 & 0.26 $\pm$ 0.105 \\ 
422 & 204.26945 & -29.82354 & 0a & 6.477 $\pm$ 0.1505 & 2.309 $\pm$ 0.0741 & 0.0 $\pm$ 0.06 \\ 
434 & 204.28178 & -29.80619 & 0a & 6.0 $\pm$ 0.1505 & 2.787 $\pm$ 0.092 & 0.71 $\pm$ 0.09 \\ 
435 & 204.27327 & -29.80574 & 0a & 6.301 $\pm$ 0.2386 & 2.289 $\pm$ 0.1106 & 0.47 $\pm$ 0.085 \\ 
436 & 204.27201 & -29.80431 & 0a & 6.602 $\pm$ 0.06247 & 2.471 $\pm$ 0.0804 & 0.0 $\pm$ 0.06 \\ 
442 & 204.26947 & -29.82319 & 0a & 6.699 $\pm$ 0.06247 & 2.511 $\pm$ 0.0847 & 0.0 $\pm$ 0.065 \\ 
443 & 204.27862 & -29.81121 & 0a & 6.602 $\pm$ 0.1109 & 2.546 $\pm$ 0.09435 & 0.0 $\pm$ 0.06 \\ 
444 & 204.26542 & -29.82648 & 0a & 6.602 $\pm$ 0.06247 & 2.593 $\pm$ 0.0894 & 0.02 $\pm$ 0.06 \\ 
456 & 204.29158 & -29.82002 & 0a & 6.477 $\pm$ 0.08805 & 2.479 $\pm$ 0.09545 & 0.35 $\pm$ 0.08 \\ 
471 & 204.28016 & -29.81992 & 0a & 6.477 $\pm$ 0.08805 & 2.343 $\pm$ 0.1236 & 0.28 $\pm$ 0.085 \\ \hline
5 & 204.26537 & -29.81034 & 0b & 6.845 $\pm$ 0.5 & 3.669 $\pm$ 0.4029 & 0.06 $\pm$ 0.04 \\ 
11 & 204.24863 & -29.81106 & 0b & 6.477 $\pm$ 0.08805 & 2.488 $\pm$ 0.1026 & 0.15 $\pm$ 0.09 \\ 
18 & 204.27221 & -29.81958 & 0b & 6.477 $\pm$ 0.06247 & 2.7 $\pm$ 0.1018 & 0.53 $\pm$ 0.065 \\ 
39 & 204.27909 & -29.82661 & 0b & 6.0 $\pm$ 0.1505 & 3.947 $\pm$ 0.0923 & 0.99 $\pm$ 0.09 \\ 
60 & 204.26685 & -29.84999 & 0b & 6.477 $\pm$ 0.1505 & 2.567 $\pm$ 0.08365 & 0.0 $\pm$ 0.055 \\ 
75 & 204.24724 & -29.85452 & 0b & 6.0 $\pm$ 0.2386 & 2.776 $\pm$ 0.1108 & 0.26 $\pm$ 0.085 \\ 
78 & 204.28363 & -29.8559 & 0b & 6.301 $\pm$ 0.2386 & 2.852 $\pm$ 0.1061 & 0.36 $\pm$ 0.085 \\ 
86 & 204.2508 & -29.86436 & 0b & 6.0 $\pm$ 0.1505 & 3.904 $\pm$ 0.1485 & 0.52 $\pm$ 0.125 \\ 
135 & 204.28913 & -29.81814 & 0b & 6.954 $\pm$ 0.07306 & 3.294 $\pm$ 0.1061 & 0.59 $\pm$ 0.035 \\ 
213 & 204.28658 & -29.83547 & 0b & 6.301 $\pm$ 0.2386 & 2.794 $\pm$ 0.1052 & 0.16 $\pm$ 0.08 \\ 
237 & 204.27946 & -29.84674 & 0b & 6.954 $\pm$ 0.1109 & 3.124 $\pm$ 0.125 & 0.7 $\pm$ 0.055 \\ 
253 & 204.2633 & -29.84839 & 0b & 6.954 $\pm$ 0.08805 & 3.938 $\pm$ 0.1003 & 0.42 $\pm$ 0.03 \\ 
298 & 204.28216 & -29.85371 & 0b & 7.176 $\pm$ 0.06735 & 3.903 $\pm$ 0.1006 & 0.34 $\pm$ 0.035 \\ 
330 & 204.28927 & -29.85945 & 0b & 6.602 $\pm$ 0.06247 & 2.964 $\pm$ 0.08645 & 0.44 $\pm$ 0.06 \\ 
336 & 204.25566 & -29.85758 & 0b & 6.301 $\pm$ 0.2386 & 4.196 $\pm$ 0.08915 & 0.79 $\pm$ 0.085 \\ 
364 & 204.25101 & -29.86449 & 0b & 6.602 $\pm$ 0.04846 & 5.13 $\pm$ 0.1349 & 1.4 $\pm$ 0.125 \\ 
387 & 204.28551 & -29.87015 & 0b & 6.0 $\pm$ 0.1505 & 3.198 $\pm$ 0.14 & 0.69 $\pm$ 0.125 \\ 
400 & 204.26281 & -29.8477 & 0b & 6.301 $\pm$ 0.2386 & 3.14 $\pm$ 0.09425 & 0.43 $\pm$ 0.08 \\ 
402 & 204.28391 & -29.85657 & 0b & 6.0 $\pm$ 0.1505 & 2.977 $\pm$ 0.1114 & 1.05 $\pm$ 0.125 \\ 
403 & 204.28286 & -29.85578 & 0b & 6.477 $\pm$ 0.08805 & 2.694 $\pm$ 0.1031 & 0.37 $\pm$ 0.075 \\ 
409 & 204.28656 & -29.85849 & 0b & 6.0 $\pm$ 0.1505 & 3.395 $\pm$ 0.0589 & 1.44 $\pm$ 0.125 \\ 
410 & 204.28834 & -29.85744 & 0b & 6.0 $\pm$ 0.1505 & 3.471 $\pm$ 0.08315 & 1.63 $\pm$ 0.125 \\ 
412 & 204.28428 & -29.86042 & 0b & 6.0 $\pm$ 0.0 & 3.082 $\pm$ 0.1373 & 1.19 $\pm$ 0.15 \\ 
419 & 204.28099 & -29.82597 & 0b & 6.477 $\pm$ 0.1505 & 2.733 $\pm$ 0.08475 & 0.64 $\pm$ 0.065 \\ 
420 & 204.27168 & -29.80613 & 0b & 6.477 $\pm$ 0.1505 & 2.606 $\pm$ 0.08815 & 0.08 $\pm$ 0.08 \\ 
421 & 204.27045 & -29.82408 & 0b & 6.0 $\pm$ 0.1505 & 2.553 $\pm$ 0.1032 & 0.45 $\pm$ 0.105 \\ 
426 & 204.29043 & -29.85845 & 0b & 6.602 $\pm$ 0.1109 & 2.473 $\pm$ 0.081 & 0.0 $\pm$ 0.06 \\ 
427 & 204.27909 & -29.83895 & 0b & 6.602 $\pm$ 0.04846 & 1.991 $\pm$ 0.1303 & 0.08 $\pm$ 0.08 \\ 
429 & 204.29133 & -29.82905 & 0b & 6.602 $\pm$ 0.1109 & 2.601 $\pm$ 0.1284 & 0.0 $\pm$ 0.075 \\ 
431 & 204.28576 & -29.82003 & 0b & 6.301 $\pm$ 0.2386 & 2.641 $\pm$ 0.08715 & 0.54 $\pm$ 0.085 \\ 
432 & 204.28776 & -29.81776 & 0b & 6.477 $\pm$ 0.1505 & 2.376 $\pm$ 0.0707 & 0.02 $\pm$ 0.065 \\ 
439 & 204.24796 & -29.81089 & 0b & 6.0 $\pm$ 0.2386 & 2.858 $\pm$ 0.0954 & 0.62 $\pm$ 0.085 \\ 
446 & 204.27934 & -29.82752 & 0b & 6.0 $\pm$ 0.1505 & 2.945 $\pm$ 0.147 & 0.87 $\pm$ 0.125 \\ 
452 & 204.25215 & -29.86739 & 0b & 6.477 $\pm$ 0.06247 & 3.364 $\pm$ 0.111 & 0.0 $\pm$ 0.065 \\ 
454 & 204.2836 & -29.84928 & 0b & 6.477 $\pm$ 0.08805 & 2.086 $\pm$ 0.0922 & 0.08 $\pm$ 0.06 \\ 
457 & 204.25226 & -29.82236 & 0b & 6.602 $\pm$ 0.04846 & 3.044 $\pm$ 0.1137 & 0.0 $\pm$ 0.075 \\ 
458 & 204.26352 & -29.82782 & 0b & 6.477 $\pm$ 0.08805 & 2.286 $\pm$ 0.1302 & 0.22 $\pm$ 0.08 \\ 
\enddata
\tablenotetext{}{This table is available for download in machine-readable format from the online version of this article.}
\end{longdeluxetable} 
\end{document}